%% file: paper.tex
\documentclass[fleqn,usenatbib]{mnras}

\usepackage{newtxtext,newtxmath}

\usepackage[T1]{fontenc}

\DeclareRobustCommand{\VAN}[3]{#2}
\let\VANthebibliography\thebibliography
\def\thebibliography{\DeclareRobustCommand{\VAN}[3]{##3}\VANthebibliography}

\usepackage{graphicx}	
\usepackage{amsmath}	
\usepackage{appendix}
\usepackage{indentfirst}
\usepackage{titlesec}
\titlespacing*{\section}{0pt}{1.1\baselineskip}{\baselineskip}

\newcommand{\ap}{$\alpha$~Per}
\newcommand{\angstrom}{\mbox{\normalfont\AA}}

\newcommand{\gaia}{{\it Gaia}}
\newcommand{\tess}{{\it TESS}}
\newcommand{\multim}{{\sc MultiModes}}
\newcommand{\mesa}{{\sc mesa}}

\input{personal_set}

\title{Determining the seismic age of the young open cluster \ap\ using $\delta$ Scuti stars }

\author[David Pamos Ortega et al.]{David Pamos Ortega$^{1}$, Antonio García Hernández$^{1}$, Juan Carlos Suárez$^{1}$, Javier Pascual Granado$^{2}$,
\newauthor Sebastià Barceló Forteza$^{1}$, José Ramón Rodón$^{2}$\\
$^1$Departamento de Física Teórica y del Cosmos, Universidad de Granada, Campus de Fuentenueva s/n, 18071, Granada, Spain\\
$^2$Instituto de Astrofísica de Andalucía (CSIC). Glorieta de la Astronomía s/n. 18008, Granada, Spain}
\date{Accepted 2022 March 24. Received 2022 March 24; in original form 2021 December 17}

\pubyear{2022}

\begin{document}
\label{firstpage}
\pagerange{\pageref{firstpage}--\pageref{lastpage}}
\maketitle

\begin{abstract}
In this work we aim at constraining the age of the young open cluster Melotte 20, known as \ap, using seismic indices. The method consists of the following steps: 1) Extract the frequency content of a sample of stars in the field of an open cluster. 2) Search for possible regularities in the frequency spectra of \dss\ candidates, using different techniques, such as the Fourier transform, the autocorrelation function, the histogram of frequency differences and the échelle diagram. 3) Constrain the age of the selected stars by both the physical parameters and seismic indices by comparing them with a grid of asteroseismic models representative of \dss. 4) Find possible common ages between these stars to determine the age of the cluster. We performed the pulsation analysis with  \multim, a rapid, accurate and powerful open-source code, which is presented in this paper. The result is that the age of \ap\ could be between 96 and 100 Myr.
This is an improvement over different techniques in the past. We therefore show that space astroseismology is capable of taking important steps in the dating of young open clusters.
\end{abstract}




\section{Introduction}\label{sec:Introduction}

\par Open clusters are laboratories of stellar and galactic astrophysics. Having at our disposal a group of tens of stars of similar chemistry and ages, within a narrow field of observation, allows us to better constrain  models with which we can compare our observations. We can better characterize the stars in this way, thereby determining their evolutionary stages and, by extension, the origin and evolution of the galaxy to which they belong.\par
There are many sources that can contribute to ambiguity when determining the age of a cluster by isochrone fitting on the Hertzsprung-Russell (HR) diagram: uncertain distance modulus,  cluster membership, binarity of individual stars, reddening and extinction, metallicity, different treatments of the physics of the models used to derive the theoretical isochrones, and the uncertainties from isochrone fitting methods. Ambiguity is even greater in young clusters, with a sparse population of stars leaving the main sequence (turn off). The tip of the red giant branch is used to date globular clusters but cannot be used in young open clusters. \par
Pulsating stars can allow a better characterization of the cluster, because some of their seismic indices are directly related to internal characteristics of the stars that we want to date. Some types of classical pulsating stars, such as \dss, have a rich and varied frequency spectra. They correspond to A and F stars of intermediate mass, very common in the main sequence (MS) for young open clusters. Space missions such as \corot\ \citep{Baglin}, \kepler\ \citep{Koch} and \tess\ \citep{Ricker} have lowered the detection threshold for their signals. Being grouped around the MS, it is not easy to estimate the age by isochrone fitting. In addition, it has been very difficult, until recently, to find regularities in the frequency spectra of these stars, as in the case of solar-type. There are studies showing that it is possible to find regularities in the frequency pattern of \dss, even out of the asymptotic regime \citep[see for example, ][and references therein]{GH2009,Paparo2016,Forteza2017,Bedding2020}. This is the named large separation, like for solar-type stars, but in the low-order regime \citep[from n=2 to n=8,][]{Suarez2014}. It is defined as the frequency difference between modes of the same degree and consecutive orders:
\begin{equation}\label{eq:dnulow}
    \fDnulow = \nu_{n,\ell} - \nu_{n,\ell-1}
\end{equation}
We used here the subscript $low$ to differentiate it from that of the asymptotic regime. However, as for solar-type stars, this \Dnulow\ is also related to the stellar mean density \citep{Suarez2014,GH2015,GH2017}.\par
Other seismic indices are the rotational splittings of modes of the same order and degree, related to the angular rotation, and the frequency at maximum power, used in solar-type stars, directly related to the effective temperature, that has also been found in \dss\ \citep[][BF2018, BF2020, BK2018 and H2021 from now on]{Forteza2018,Forteza2020,Bowman2018,Hasanzadeh2021}.\par
Here we use the asteroseismic indices to  study the age of the open cluster Melotte 20 (also known as \ap). The structure of the paper is as follows: in Sec.~\ref{sec:alphaPer} we provide the main characteristics of the cluster and former estimates of its age. In section Sec.~\ref{sec:Data} we describe the sample of stars used in this research. In Sec.~\ref{sec:analysis} we introduce the code \multim, a new tool for the analysis of pulsating stars. In order to show its reliability, we have compared it to one of the most reliable and most used in the field, \sigspec\ \citep{Reegen2007}. We explain the details of the code in this section, and also the comparative results obtained between them, using synthetic and real light curves, in terms of accuracy and computing time. In this section we also present the list of \dss\ candidates identified in our sample with the frequency analysis. In Sec.~\ref{sec:Regularities} we show how, in four of them, it has been possible to measure the low-order large separation. The angular rotation and the frequency at maximum power of two of these stars have been measured. In Sec.~\ref{sec:Grid} we explain the details of the grid of pulsation models calculated with \mesa\ \citep{Paxton2019} and \filou\ \citep{SG2008}, in order to constrain the models that better fit to the seismic observations. In Sec.~\ref{sec:Age} we compare our asteroseismic age with previous works and discuss the reliability of the proposed method to date open clusters. Finally, in Sec.~\ref{sec:Conclusions}, the main conclusions are exposed.
\vspace{-1ex}

\section{\texorpdfstring{$\boldsymbol{\alpha\ P\MakeLowercase{er}}$}{alpha Per}}\label{sec:alphaPer}

The open cluster \ap\ is located in the constellation of Perseus, at a distance of $174.89\pm0.16$ pc, obtained from \gaia\ DR2 parallaxes \citep{GaiaCollaboration2018a}. In \citet{Lodieu2019} a total of 517 astrometric member candidates had been identified in the tidal radius of \ap, using a kinematic method combined with the statistical treatment of parallaxes and proper motions. \citet{Netopil2013} estimated solar metallicity for all members of the cluster and an extinction along the line of sight of around $A_{V} = 0.3$ \citep{Prosser1992}. \par
Recent age estimates by isochrone fitting range from 20 to 90 Myr. \citet{Makarov2006} estimates the age in around 52 Myr, fitting isochrones in the $M_{V}$ vs. (B-V) diagram, with Z = 0.02, E(B-V) = 0.055, and overshooting computed from the models by \citet{Pietrinferni2004}. According to \citet{Silaj2014}, the age of \ap\ is $60\pm7$ Myr, determined with isochrones by \citet{Girardi2000}, with Z = 0.02, fitted onto an HR diagram, considering that the uncertainties are less ambiguous than using a colour-magnitude diagram. Isochrones are most constrained around the one that passes throw the brightest main sequence star, $\psi$ Per (HD 22192), which lies close to the terminal age main sequence (TAMS). In both works, the estimated age of the cluster relies on the parameters of a single star, with a difference of around 8 Myr in their results. Such an approach must be taken with caution, since errors in the estimate of luminosities and temperatures for stars in different evolutionary stages can be quite large. These errors may come from, e.g. the type of photometry to use, bolometric corrections and the correct modulus distance, or even the lack of corrections for rotation effects \citep[see e.g.][ for more details]{suarez2002}. \par
Other works use spectroscopic observations of lithium in low-mass stars to estimate the age of young clusters (see for example \citet{Basri1999}, \citet{Stauffer1999}). The idea is that for stars near the substellar mass limit, the age is very sensitive to the lithium depletion boundary, when they start to burn lithium in their cores. As these stars are fully convective in their range of masses, the abundance of lithium is visible in their atmospheres through the 6708 \angstrom\ Li I doublet. The problem with this method is that unresolved binaries makes the cluster appear younger, as claimed by \citet{Martin2001}. For this reason, they used the faintest members to derive the age of the cluster, for which lithium has not been detected. Another difficulty with this method is its dependency on good models that take into account the characteristics of the stellar atmosphere. In \citet{Stauffer1999} the estimated age with this method is around $90 \pm 10 Myr$, noticeably older than \citeauthor{Makarov2006} and \citeauthor{Silaj2014}. \par
\citeauthor{Lodieu2019} assumes an age of 90 Myr to compile the census of stars that are members of the cluster,  because it is the most common age in the literature.\par

\section{Our sample}\label{sec:Data}

\begin{figure*}   
    \centering
    \includegraphics[width=0.9\textwidth]{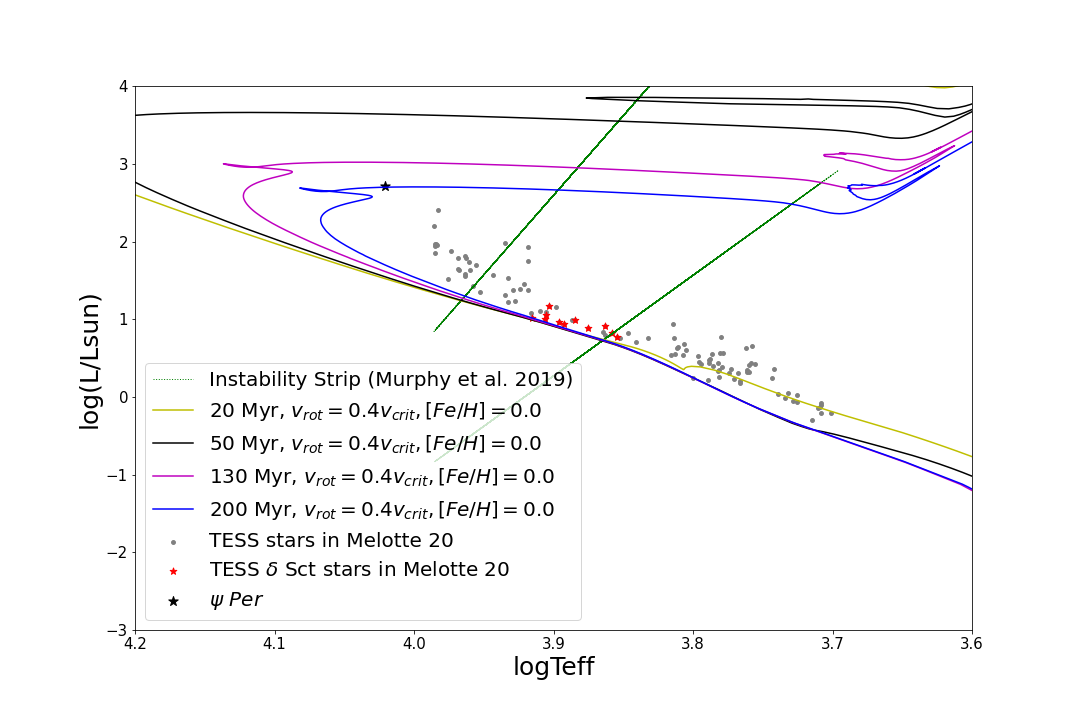}
    \caption{HR for the sample of 112 stars from the field of $\alpha$ Per, taken from TESS Input Catalogue. Isochrones between 20 and 200 Myr with solar metallicity are taken from MIST. (\textcolor{blue}{\href{http://waps.cfa.harvard.edu/MIST/}{http://waps.cfa.harvard.edu/MIST/}}). The borders of the instability strip are also drawn, calculated according to \citet{Murphy2019}. The position of $\psi$ Per is also indicated, one of the most luminous stars in the cluster}
    \label{fig:HR}
\end{figure*}

Firstly, we have done a cross match between VizieR Online Data Catalogue Gaia DR1 open cluster members \citep{GaiaCollaboration2017} and TESS Input Catalogue (TIC) \citep{Stassun2019}, searching possible targets belonging to the same open cluster, from which we could obtain light curves from the TESS mission. We have found a list of 112 stars in the field of \ap, with measured values for parallax, mean G magnitude, E($B-V$) reddening, bolometric corrections estimated by \citet{Andrae2018}, and effective temperature. They are all contained in the census prepared by \citeauthor{Lodieu2019}, obtained from \citet{GaiaCollaboration2018a}. We have estimated the G magnitude with the equation:
\begin{equation}\label{eq:Gmag}
    M_{G} = G - 5\log r + 5 - A_{G},
\end{equation}
where the distance $r$ is taken as the inverse of the \gaia\ parallax. The extinction $A_{G}$ is calculated from reddening values as in \citeauthor{Stassun2019}:
\begin{equation}\label{eq:Extinction}
    A_{G} = 2.72\,\mathrm{E}(B-V).
\end{equation}
The G magnitude is converted to luminosity using the bolometric correction BC(T\textsubscript{eff}) with the equation:
\begin{equation}\label{eq:Luminosity}
    -2.5\log L = M\textsubscript{G} + \mathrm{BC}\textsubscript{G}(T\textsubscript{eff}) - M_{\mathrm{bol}\odot}
\end{equation}
Figure~\ref{fig:HR} depicts the HR diagram of our sample, including MIST\footnote{\href{http://waps.cfa.harvard.edu/MIST/}{http://waps.cfa.harvard.edu/MIST/}} isochrones \citep{Dotter2016,Choi2016,Paxton2011,Paxton2013,Paxton2015}, with solar metallicity and different ages. They all have been computed from the pre-main sequence phase (PMS) to the end of hydrogen burning. In rotating models, solid-body rotation with $\Omega / \Omega_{crit} = 0.4$ is initialized at zero age mean sequence (ZAMS). Rotation makes the star hotter or cooler depending on the efficiency of rotational mixing in the envelope. If rotational mixing introduces a sufficient amount of helium into the envelope, increasing the mean molecular weight, the star becomes more compact and hotter. On the contrary, if rotational mixing is not efficient, the centrifugal effect dominates, making the star cooler and more extended. The ages of the isochrones are taken to cover the wide range of values found in the literature (see previous section). We can see that all the calculated isochrones between 20 and 200 Mrs mostly overlap in the MS zone. We also have located the position of $\psi$ Per on the HR diagram, which is compatible with an age of 200 Myr, quite far from the age estimated by \citeauthor{Silaj2014}. \par
This is where asteroseismology plays an important role, since it provides independent seismic indices that provide an accurate determination of global stellar parameters, including the age of the star.\par
\begin{figure}   
    \centering
    \includegraphics[width=0.45 \textwidth]{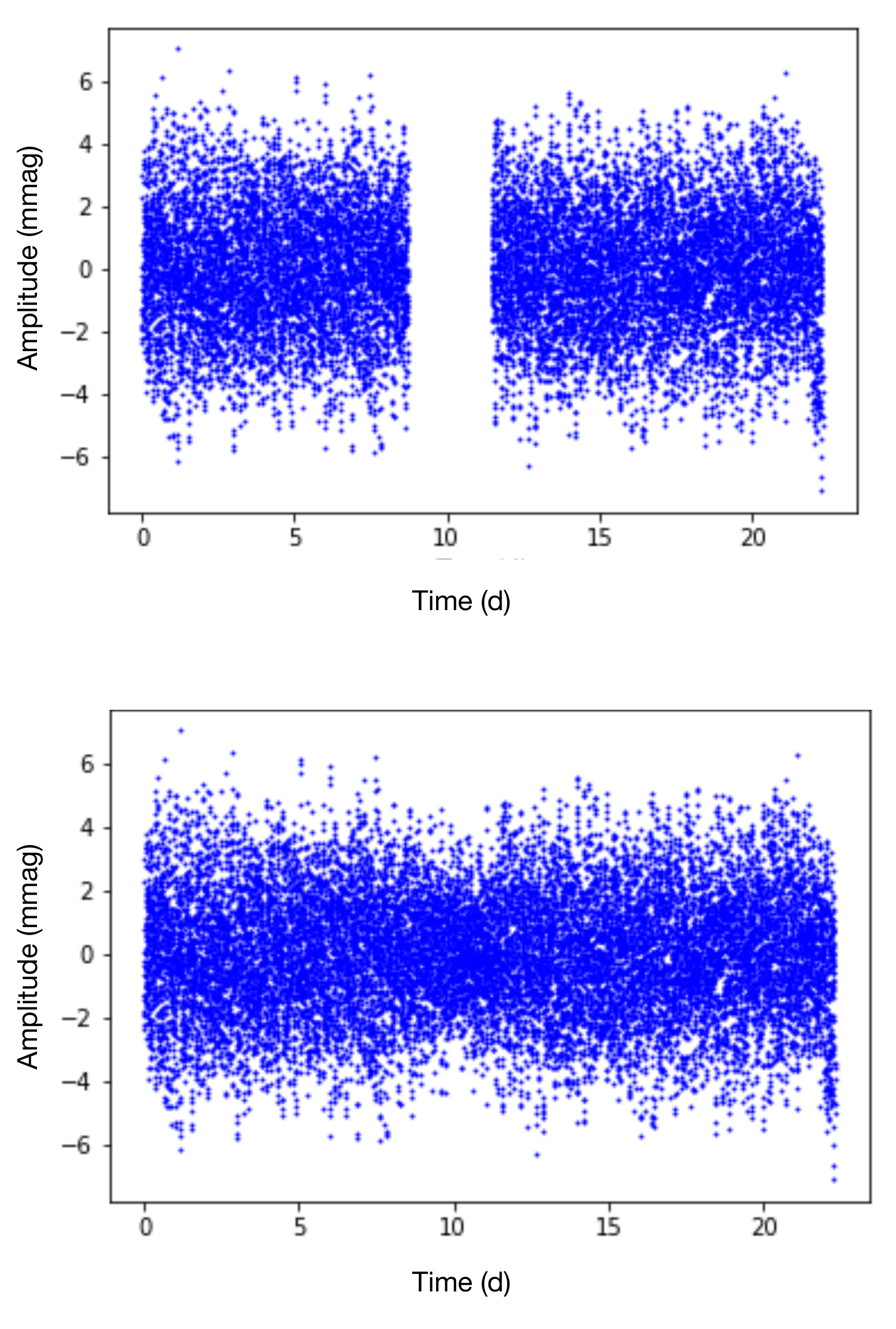}
    \caption{Top: Light curve of TIC 252851046. Bottom: Interpolated light curve of TIC 25851046 with MIARMA algorithm}
    \label{fig:Lightcurves}
\end{figure}

We analyzed a set of 32 stars using data from sector 18 of the \tess\ mission, with approximately 14700 points, a Rayleigh resolution of approximately 0.045 $d^{-1}$ and a cadence of two minutes. We used the Pre-Search Data Conditioned (PDC) light curves, corrected for instrumental effects, that are publicly available through the  MAST\footnote{\href{https://archive.stsci.edu/}{https://archive.stsci.edu/}}. These light curves have a 2.5 days gap caused by the satellite downlink, when the spacecraft passed through the shadow of the Earth at the start of orbit 43 (Fig.~\ref{fig:Lightcurves}). \citet{Pascual-Granado2018} showed that the analysis of light curves with gaps can produce an undesirable amount of spurious peaks. Following \citet{Pascual-Granado2015} we interpolated our data using the {\sc MIARMA} code, which uses autoregressive and moving average models. This method is aimed to preserve the original frequency content of the light curve, making the frequency extraction more reliable than leaving the gaps or using other gap-filling methods.

\section{Frequency analysis}\label{sec:analysis}

\subsection{MultiModes: a new tool for analysis of pulsating stars}\label{subsec:multimodes}

Many of the light curves provided by space missions are uniformly sampled, but show some gaps, due to lack of observations, instrumental issues or environmental effects. The Lomb Scargle Periodogram (from now on LS) is very powerful for analyzing non-uniformly sampled time series \citep{Scargle}. 
For uniform sampling, LS provides the classical periodogram. The algorithm calculates, at each frequency of interest, a time phase shift, and which in turn is used to evaluate the power spectra at that frequency. This time phase shift makes the LS to be independent of shifting all the points by any constant. In this way, the calculation of LS is equivalent to performing a least squares fit of data to a sinusoidal function for each evaluated frequency. So there is a deep connection between the Fourier Transform and least squares analysis in the LS. \par

Computing LS can be very slow because it requires a number of calculations of the order of $N^2$, $N$ being the number of points in the sample. This problem is partially solved by the implementation of the so-called Fast Lomb Scargle Evaluation \citep{Press}, based on \textit{extirpolation}\footnote{Extirpolation consists of the substitution of actual points of the series by others constructed by means of an evenly sampled mesh. On this basis, the order of the calculations for the evaluation of the periodogram is reduced to $N\log N$} from unevenly sampled data.\par

The width of a peak does not depend on the number of sampled points or the signal to noise ratio (SNR), it only depends on the Rayleigh resolution, which is inversely proportional to the total sample size \citep{Vanderplas}. Two peaks within the Rayleigh resolution, in theory, could not be perfectly resolved by the periodogram. Furthermore, the stochastic nature of the series can cause numerous spurious frequencies to emerge, many of them of low amplitude but above the mean noise level. As detailed in  \citet{Balona2014}, the problem of using the pre-whitening technique lies in the large number of artificial frequencies that are generated with a high SNR. When, at each step, a signal very similar to the one to be extracted is added to the analyzed light curve, if the fit has not enough accuracy, it can generate an interference pattern around the added frequency. \par

These spurious frequencies can be avoided, to a great extent, by making a simultaneous fit to all the extracted frequencies, through non-linear optimization. We use least squares fit to a multisine function, taking as parameters the frequencies, amplitudes and phases of each of the peaks to be extracted. \par
The well-known \sigspec\ algorithm \citep[][SS from now on]{Reegen2007} uses pre-whitening to extract all those frequencies that are above a certain criterion of significance, defined from the False Alarm Probability (FAP) as:
\begin{equation}\label{eq:sig}
    \mathrm{sig} = -\log (\mathrm{FAP})
\end{equation}
taking into account the dependency of the FAP with the phases.\par 
Widely used in asteroseismology, it is not open source, and no longer updated.\par
Inspired by the \sigspec\ methodology, we developed \multim\ (MM from now on), which is a Python routine developed to extract the most significant peaks of a sample of classical pulsating stars. It is available in a public Github repository\footnote{\href{https://github.com/davidpamos/MultiModes}{https://github.com/davidpamos/MultiModes}} and it is customizable. This routine has been designed using the Astropy package \footnote{\textcolor{blue}{\href{https://www.astropy.org}{https://www.astropy.org}}} to calculate the periodograms and the LMFIT package \footnote{\textcolor{blue}{\href{https://lmfit.github.io/lmfit-py/}{https://lmfit.github.io/lmfit-py/}}} for non-linear optimization of the extracted signals.\par

\begin{figure*}   
    \centering
    \includegraphics[width=0.9 \textwidth]{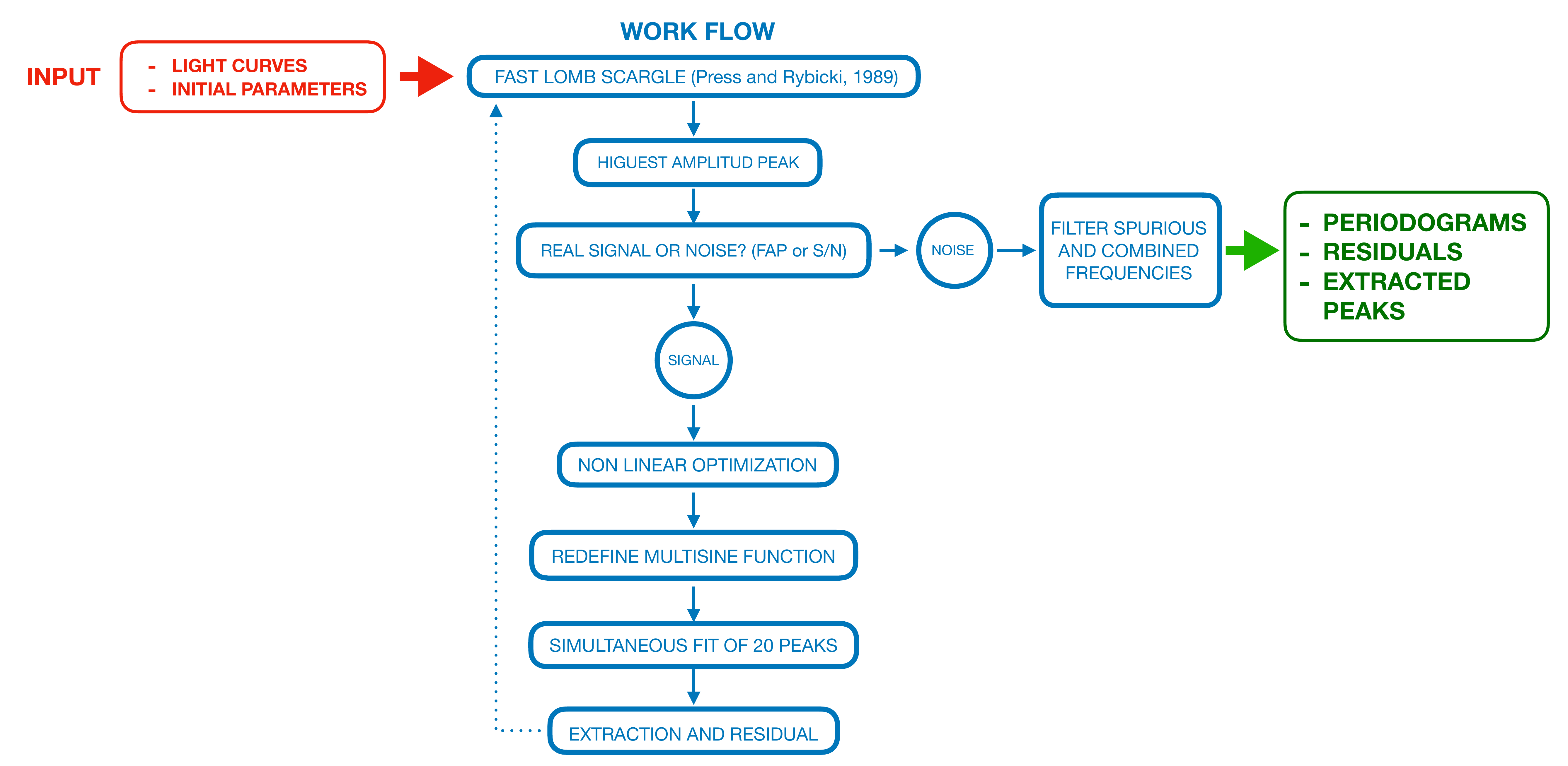}
    \caption{Workflow of the MultiModes algorithm}
    \label{fig:Workflow}
\end{figure*}

The algorithm (see Fig.~\ref{fig:Workflow}) fits frequency, amplitude and phase through non-linear optimization, using a multisine function. This function is redefined with the new calculated parameters. It does a simultaneous fit of a number of sinusoidal components (20 are usually enough).\par
Then they are subtracted from the original signal and the algorithm goes back to the beginning of the loop with the residuals as input, repeating the same process, until the stop criterion is reached. After that, the code can filter suspicious spurious frequencies, those of low amplitude above the Rayleigh resolution, and possible combinations of modes.
\vspace{-1ex}

\subsection{Accuracy test with simulated spectra}\label{accuracy}

We tested the reliability of MM using synthetic light curves. Following \citet{Balona2014}, we constructed artificial light curves with 50, 100, 200 and 400 frequencies, with uniformly distributed values between 0 and 30 d$^{-1}$, amplitudes ranging between 0 and 10 mmag, exponentially distributed towards low values, phases uniformly distributed, and adding a Gaussian noise of about 0.5 mmag. 

\begin{figure*}   
    \centering
    \includegraphics[width=0.8 \textwidth]{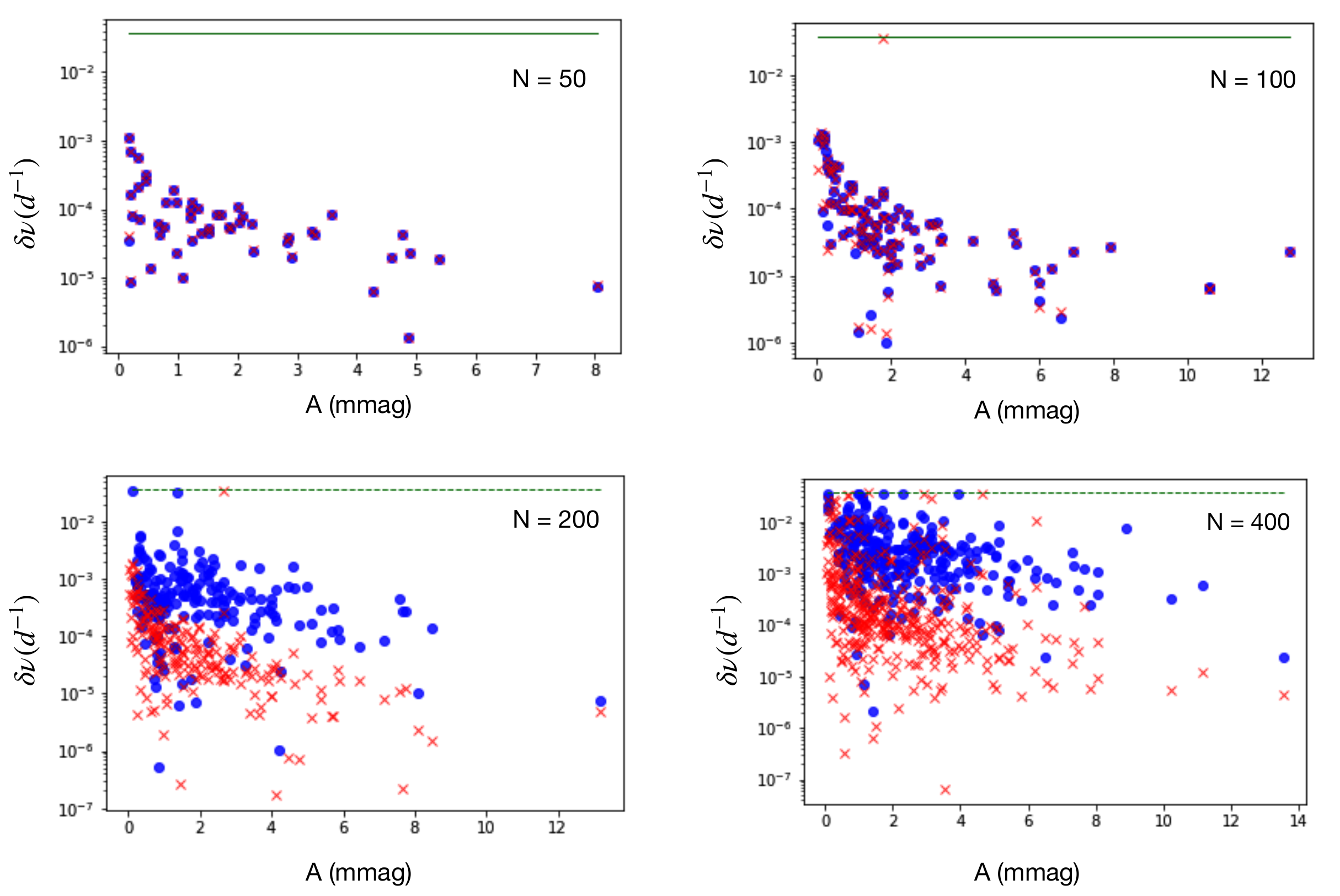}
    \caption{Comparative analysis of accuracy between MM and SS, using simulated light curves with different number of frequencies, 50 (upper left panel), 100 (upper right panel), 200 (bottom left panel) and 400 (bottom right panel). Each plot represents the frequency deviation of the extracted peaks with MM (blue filled circles) and SS (red crosses). The dotted green line is the level of the Rayleigh resolution.}
    \label{fig:MMvsSSaccuracy}
\end{figure*}

The frequency deviation of MM is similar to that of SS in the curves with 50 and 100 frequencies, around $10^{-4}\ d^{-1}$, when we used an unlimited number of components for the simultaneous fit of the simulated signal (Fig.~\ref{fig:MMvsSSaccuracy}). With a higher number of frequencies, 200 and 400, SS is more accurate than MM by one order of magnitude, because MM used a limited number of components for the simultaneous fit, no more than 50. Still, results with MM have a high enough accuracy for a precise frequency extraction. As the frequency density increases, the problem becomes more unstable since more spurious frequencies appear, with both MM and SS. Therefore, it makes sense to limit the number of components of the simultaneous fit of the light curve when frequency density is very high. In this sense, MM allows us to have greater control over the frequency analysis, by limiting the number of components for the fit, and working with frequency packs.
 
\subsection{\texorpdfstring{Frequency content of the $\boldsymbol{\delta}$ Scuti sample}{Frequency content of the delta Scuti sample}
}\label{subsec:Frequencies}

From the sample of 32 analysed stars analyzed with MM, we found that 11 of them are \dss, or hybrids, a type of pulsating stars with intermediate mass that show low-order acoustic oscillations (p-modes) and also high order gravity modes (g-modes), more typical of $\gamma$ Dor stars \citep{Grigahcene2010,Uytterhoeven2011}. Except for TIC 252829836 and TIC 347570557, 9 stars in our sample had not been identified as \dss\ so far. \par
We have analysed the frequency content of these 11 stars using MM and SS, for greater reliability. The frequency comparison assumes that both codes identify the same frequency when their difference is smaller than the Rayleigh resolution. We established the degree or percentage of coincidence based on whether each frequency extracted by MM is among those extracted by SS or not. The degree of coincidence between MM and SS in the values of the extracted frequencies, extracting the most significant frequencies, is above $90\%$ in all the cases (Tab.~\ref{tb:MMvsSSfreqs}). We have also done a comparative analysis on the computing speed between MM and SS with our sample of 11 \dss\ stars (see Ap.~\ref{appendixA}). 

    \begin{table}
	\centering
	\caption{Comparative analysis between SS and MM extracted peaks, using  ARMA-interpolated light curves with {\sc MIARMA} code, for the 11 \dss\ sample from \ap. $N$\textsubscript{SSar} and $N$\textsubscript{MMar} are, respectively, the number of extracted frequencies by \sigspec\ and \multim. SSarMMar is the degree of coincidence between \sigspec\ and \multim}
	\renewcommand{\arraystretch}{2}
	\addtolength{\tabcolsep}{2 pt}
	\resizebox{8.5 cm}{!}{
    \begin{tabular}{ccccc}
			\hline
			TIC &  $N$\textsubscript{SSar} & $N$\textsubscript{MMar} & SSarMMar\\
			\hline
			104319359 &  249 & 238  &  97.9\\
			116011834 &  233 & 222   & 97.7\\
			252829836 &  82 & 76 &  97.4\\
			252851046 &  213 & 222  & 93.7\\
			285935852 &  45 & 40  & 97.5\\
			347570557 &  189 & 188  & 94.7\\
			354638295 &  97 & 89  & 98.9\\
			354792288 &  55 & 47  & 100.0\\
			401079326 &  39 & 35  & 100.0\\
			410732825 &  69 & 55   & 100.0\\
			428320122 &  71 & 64   & 100.0\\
    \end{tabular}
    }
	\label{tb:MMvsSSfreqs}
    \end{table}

\section{\texorpdfstring{Seismic indices of $\boldsymbol{\delta\ S\MakeLowercase{cuti}}$ stars}{Seismic indices of delta Scuti stars}}\label{sec:Regularities}

\begin{figure*}   
    \centering
    \includegraphics[width=0.8\textwidth]{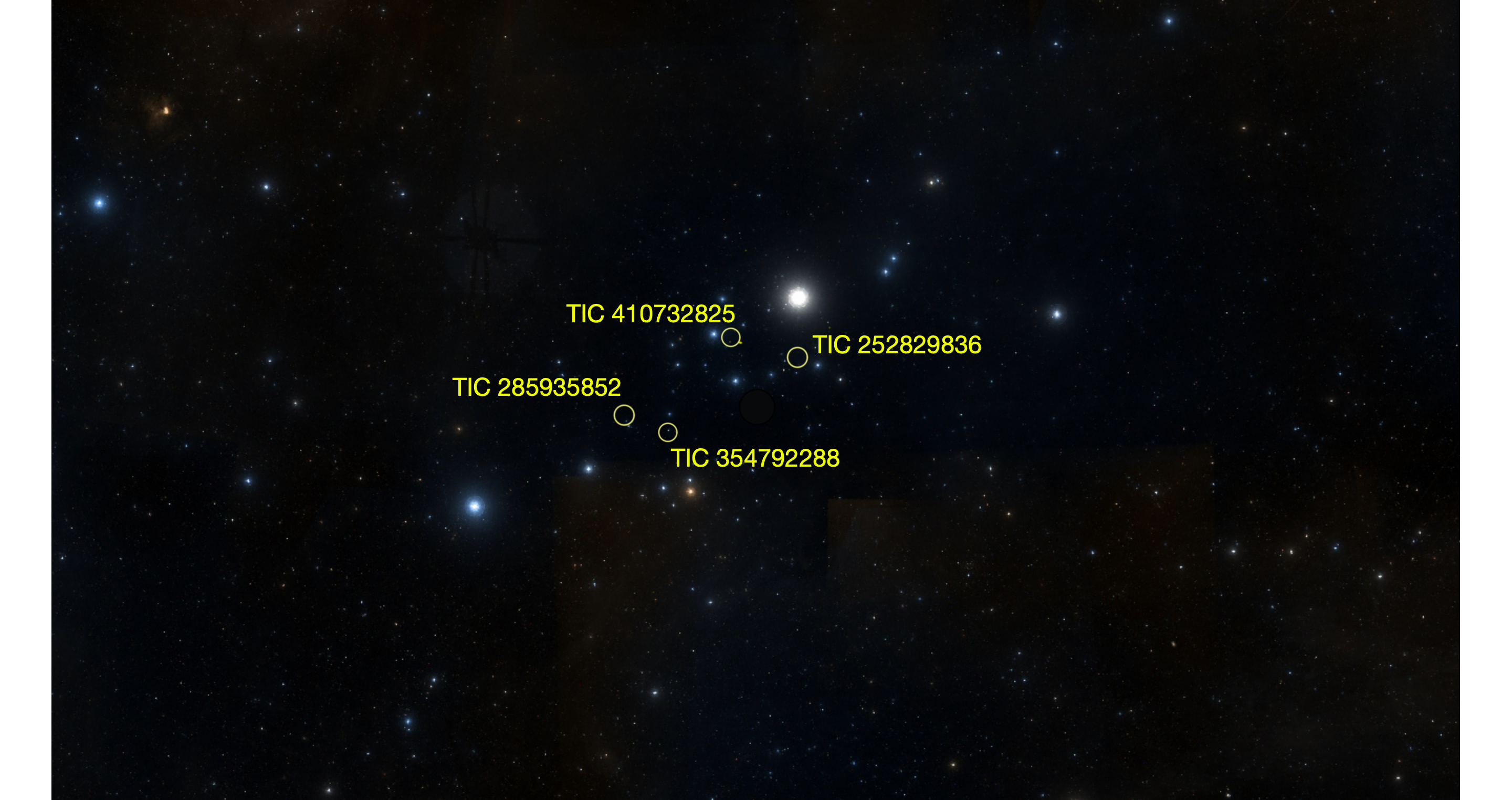}
    \caption{The four selected targets in the field of \ap\, for which we have obtained regularities in the frequency spectra}
    \label{fig:alphaPertargets}
\end{figure*}

Among our sample of 11 \dss, in four of them, TIC 410732825, TIC 354792288, TIC 285935852 and TIC 252829836 (Fig.~\ref{fig:alphaPertargets}), we found a pattern that we identify as a low-order large separation ($\fDnulow$), following the techniques from \citet{GH2009} and \citet{RB2021}. The autocorrelation function (AC), the Fourier transform (FT), the histogram of frequency differences (HFD), and the échelle diagram (ED) were applied to the 30 frequencies with highest amplitudes and above 5 \cd\ to avoid g~modes. These frequencies are selected by amplitude but they are given equal amplitudes when computing the transformations. \par
\begin{figure*}
    \centering
    \includegraphics[width=1.0 \textwidth]{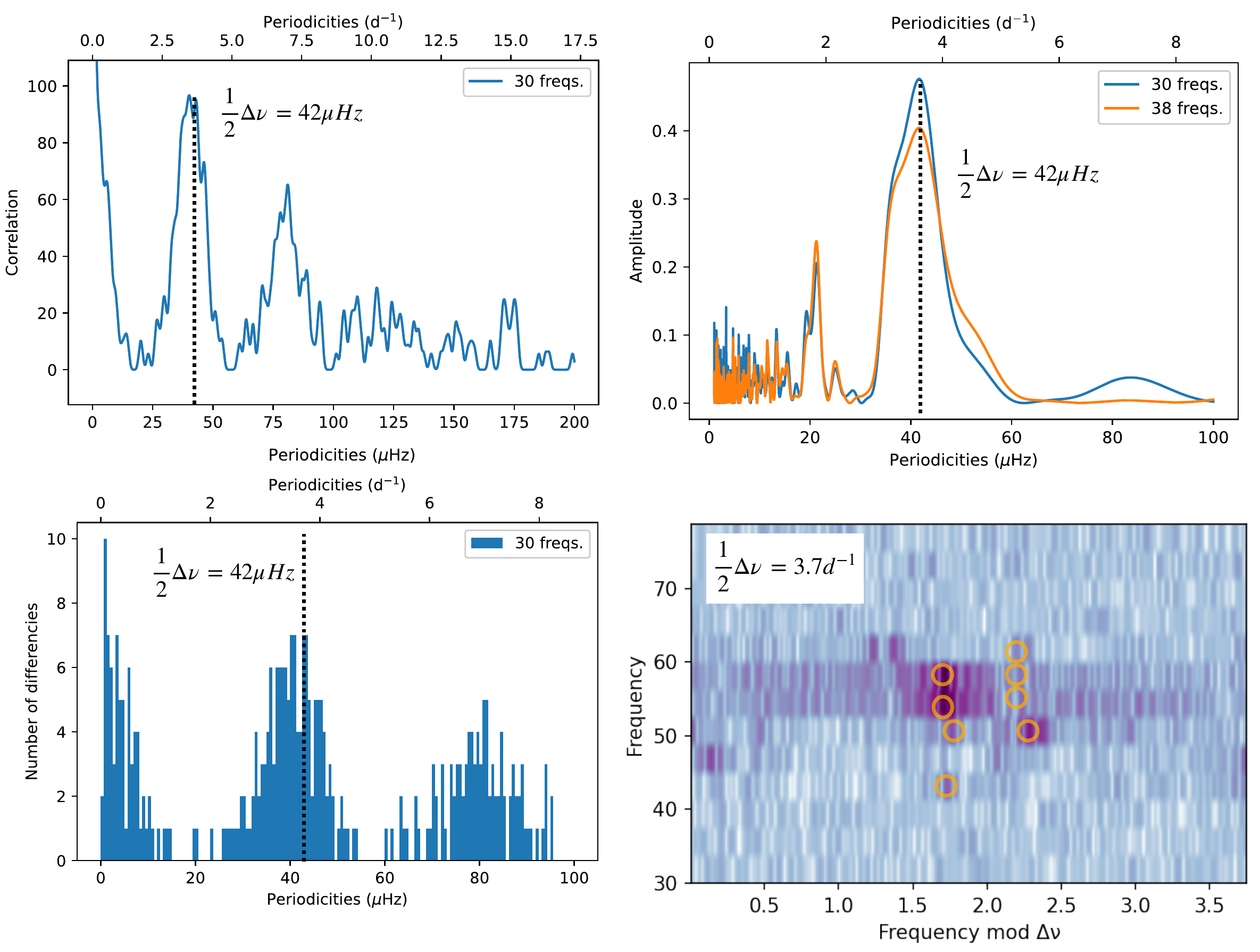}
    \caption[Caption for LOF]{Measured regularities in the frequency spectra of TIC 354792288. The top left panel represents the autocorrelation diagram (AC) of periodicities calculated with the first 30 frequencies extracted by MM, showing peaks around 42 $\mu$Hz (black dotted line) and 84 $\mu$Hz. Top right, the Fourier transform (FT) for the periodicities of those first 30 frequencies, showing a peak around 42 $\mu$Hz (black dotted line). Bottom left, the histogram of frequency differences (HFD) with the first 30 frequencies, showing peaks around 42 $\mu$Hz (black dotted line) and 84 $\mu$Hz. Bottom right, the echelle diagram (ED) showing two vertical ridges (orange circles) when 1/2$\fDnulow$ is chosen around 3.7 $d^{-1}$, equivalent to 43 $\mu$Hz. ED is done with Python package Echelle 1.5.1, developed by \citeauthor{Hey2020} \citeyear{Hey2020}\protect\footnotemark[3]}
    \label{fig:TIC88}
    
\end{figure*}
\begin{figure*}   
    \centering
    \includegraphics[width=1.0 \textwidth]{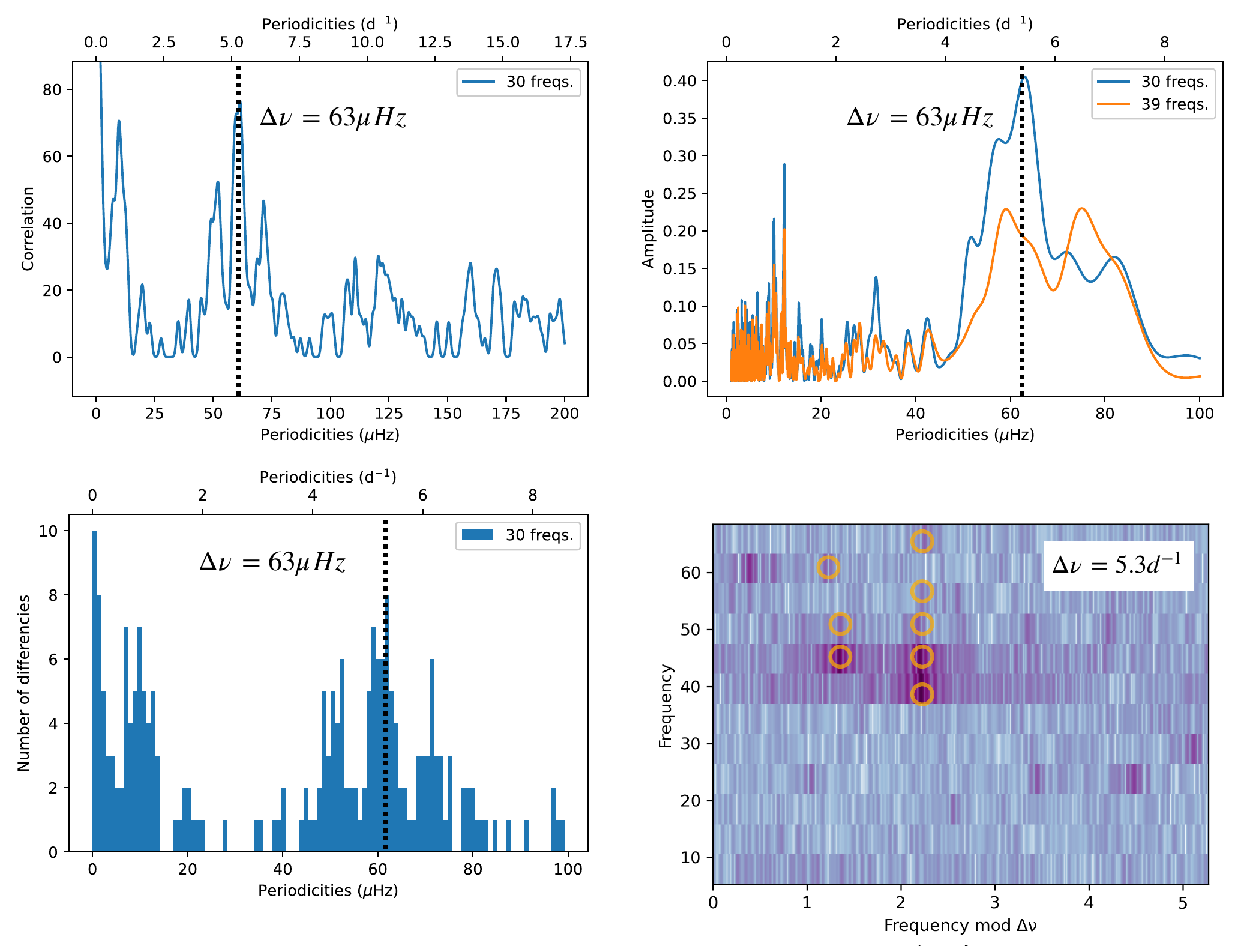}
    \caption{Measured regularities in the frequency spectra of TIC 410732825. The autocorrelation diagram (AC) (top left), the Fourier transform (FT) (top right), and the histogram of frequency differences (HFD) (bottom left), calculated with the first 30 frequencies extracted by MM, show a prominent peak around 63 $\mu$Hz (black dotted line). And also, the echelle diagram (ED) (bottom right) show a very clear vertical ridge (orange circles) when $\fDnulow$ is around 5.3 $d^{-1}$, equivalent to 61 $\mu$Hz}
    \label{fig:TIC25}
\end{figure*}

Fig.~\ref{fig:TIC88} and Fig.~\ref{fig:TIC25} show, respectively, the cases of TIC 354792288 and TIC 410732825 (see the other two stars in Ap.~\ref{appendixB}). For TIC 354792288, the FT, AC and HFD show a peak around $42\ \mu \mathrm{Hz}$. It may be half the value of the low-order large separation. The AC and HFD also show a peak around $84\ \mu \mathrm{Hz}$. The ED shows the alignment of several frequencies when 86~\muhz\ is chosen as the low-order large separation. Thus, we took 84~\muhz\ as \Dnulow, being the most common value obtained by the different diagnostic techniques.\par
We also searched for a pattern connected to the rotational splitting in the p-mode regime. At first order in the perturbative theory, rotation splits the oscillation modes of the same order and spherical degree in the form:
\begin{equation}
    \label{eq:splitting}
    \omega_{nlm} = \omega_{nl} + m\Omega(1 - C_{L}),
\end{equation}
where $\omega$ is the angular frequency, $\Omega$ is the angular rotation frequency and $C_{L}$ is the Ledoux constant, to take into account the effects of the Coriolis force \citep{Aerts2010}. According to Eq.~\ref{eq:splitting}, this would allow us to find a pattern corresponding to the rotation frequency in the periodogram, i.e., a regular structure related to $\Omega$. However, this simple distribution of frequencies is only valid for slow rotators. Even at moderate rotations (around 50~\kms) the picture changes: rotational splittings are not symmetric anymore and even the m = 0 mode can be displaced \citep[see e.g. ][ for more details]{suarez2006}. These effects may hamper the identification of a rotational splitting.\par

Nonetheless, some theoretical works computing 2D non-pertubative models and rapid rotation \citep{Reese2017} combined the AC and the FT \citep[following][]{GH2009} to search for the patterns of the low-order large frequency separation and the rotational splitting. They pointed out that \Dnulow\ or its half value comes up usually clearer in the FT, whereas twice the rotational splitting can be better found with AC diagnostic. \par
From the observational side, \citet{Paparo2016} searched for regularities in the frequency spectra of \dss\ observed by \corot\ using a visual inspection and by means of a semi-automatic algorithm that looks for repeated spacings between frequencies. They claimed that some of these spacings might be not only the large separation but also the rotational splitting or a linear combination of both. The work of \citet{Forteza2017} used the AC, HFD, ED and DFT of the power spectrum, following the techniques from \citet{Regulo2002}, to find rotational splittings in a sample of four \dss. The main idea was that structural characteristics of the star, such as rotation, alter the power spectra, particularly in the flat plateau, where the density of low amplitude frequencies increases with the rotation rate. \citet{RB2021} related regularities found in the frequency spectra of binary \dss\ to their rotational splitting. In their work, they used the FT, the AC and the HFD as in the present research.\par
The HFD for two of our stars, TIC 410732825 (Fig.~\ref{fig:TIC25}) and TIC 285935852 (Fig.~\ref{figa2}) shows a prominent peak around 9-11 \muhz\ that could be the identified as the angular rotation. In fact, Fig.~\ref{fig:TIC25} also shows significant peaks at both sides of the one identified as \Dnulow, implying the presence of spacings corresponding to the large separation plus and minus the rotational splitting.\par
\begin{figure*}   
    \centering
    \includegraphics[width=1.0 \textwidth]{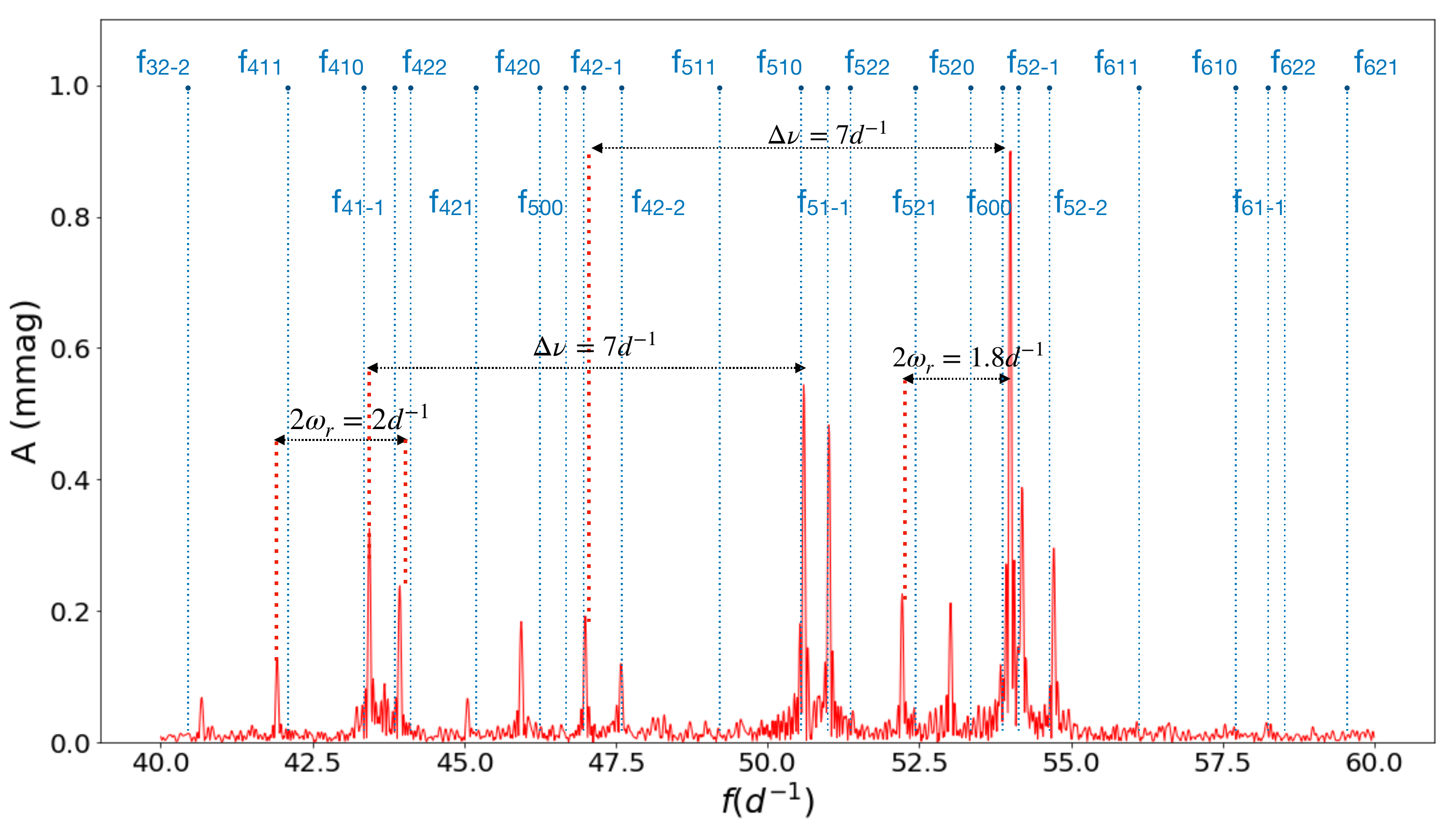}
    \caption{Periodogram of TIC 285935852, showing a measured low-order large separation of around $\fDnulow = 7 d^{-1}$, and a rotational splitting of around \textcolor{red}{2}$\omega_{r} = 1.9 d^{-1}$. The modes distribution is compared to a suitable model calculated with \mesa-\filou (see Sec.~\ref{sec:Grid}). The blue dotted lines show the positions of some of the theoretical modes, denoted as $f_{nlm}$, where \textit{n} is the corresponding radial order, \textit{l} is the spherical degree and \textit{m} the azimutal order}. Some modes seem to follow the frequency distribution of the model. Others, related to rotation, where the frequency splitting is visible, don't do it. Even the centroids are displaced from the model
    \label{fig:TIC52vsmodel}
\end{figure*}

In the case of TIC 285935852, Fig.~\ref{fig:TIC52vsmodel} displays the periodogram with the most visible modes, between 40 and 60 $\mu Hz$. It is indicated the low-order large separation, with a value of around $\fDnulow = 7 d^{-1}$. Some other periodicity can also be noticed that might correspond to twice the rotational splitting of around 0.9-1.0 $d^{-1}$, or equivalently, 9-11 \muhz. This is consistent with the results by \citet{Reese2017}. The $m\neq0$ modes are not symmetric with respect to the $m=0$ mode, that it is displaced but a spacing corresponding to double the rotational splitting still remains. To check this behaviour, the mode distribution is compared to a compatible model and set of pulsation frequencies computed with the \mesa\ and \filou\ codes (see Sec.~\ref{sec:Grid}), taking rotation into account up to second order in the pertubative theory for the adiabatic oscillation computation (including near-degeneracy effects and stellar structure deformation). The mode distribution is similar in both the model and the observation although the limitations of our calculations are clear. Some theoretical modes lie close to the observed ones but others do not. There are missing or completely off modes from the model. Fortunately, the spacings remain although our mode identification might mismatch if we use non-rotating models. We emphasise the importance of taking rotation into account when modelling \dss, as also pointed out by \citet{Murphy2021}, although they used non-rotating models to carry out a mode identification.\par
Another seismic index, recently discovered for \dss, is the frequency at maximum power, related to the effective temperature of the star, as shown in the works of BF2018, BF2020, BK2018 and H2021. 
There are different ways of defining the frequency at maximum power. BK2018 defines it with the peak of maximum amplitude in the spectrum, $\nu$\textsubscript{max,0}. BF2018 and BF2020 do it in two different ways, one of them with $\nu$\textsubscript{max,0} and the other through the mean of the most significant extracted peaks weighted by their amplitudes, $\nu$\textsubscript{max,w}.  H2021 do it through the maximum of the autocorrelation of the power spectra, $\nu$\textsubscript{max,2D}. As is explained in \citet{Hasanzadeh2021}, and with more detail in \citet{Viani2019}, the AC method is applied on several windows of the SNR periodogram, to obtain the range of frequencies for the modes' envelope. Then, a Gaussian curve is fitted to the mean collapsed correlation with each window. The peak of this curve is $\nu$\textsubscript{max,2D}. \par 
This relation is very dependent on the evolutionary stage of the star through \textit{log g}, and it is affected by gravity darkening, as a result of the rapid rotation. This quite large dispersion is probably intrinsic to the relationship itself, and also due to observational errors. The relation probably does not have the same origin as for solar-like stars since the excitation mechanism of the modes is very different in these two types of pulsators. \par
Tab.~\ref{tb:indices} shows the values of all these seismic indices that will serve to constrain the models to try to date these four stars accurately, and therefore, the cluster. The uncertainty of $\nu$\textsubscript{max,0} is the error of the single frequency extracted with MM. The uncertainty in the location of $\nu$\textsubscript{max,w} has been calculated from the standard deviation of the amplitude weighted data. For $\nu$\textsubscript{max,2D}, we have taken the standard deviation of the Gaussian fit  as its corresponding error.\par
\begin{figure*}   
    \centering
    \includegraphics[width=0.9 \textwidth]{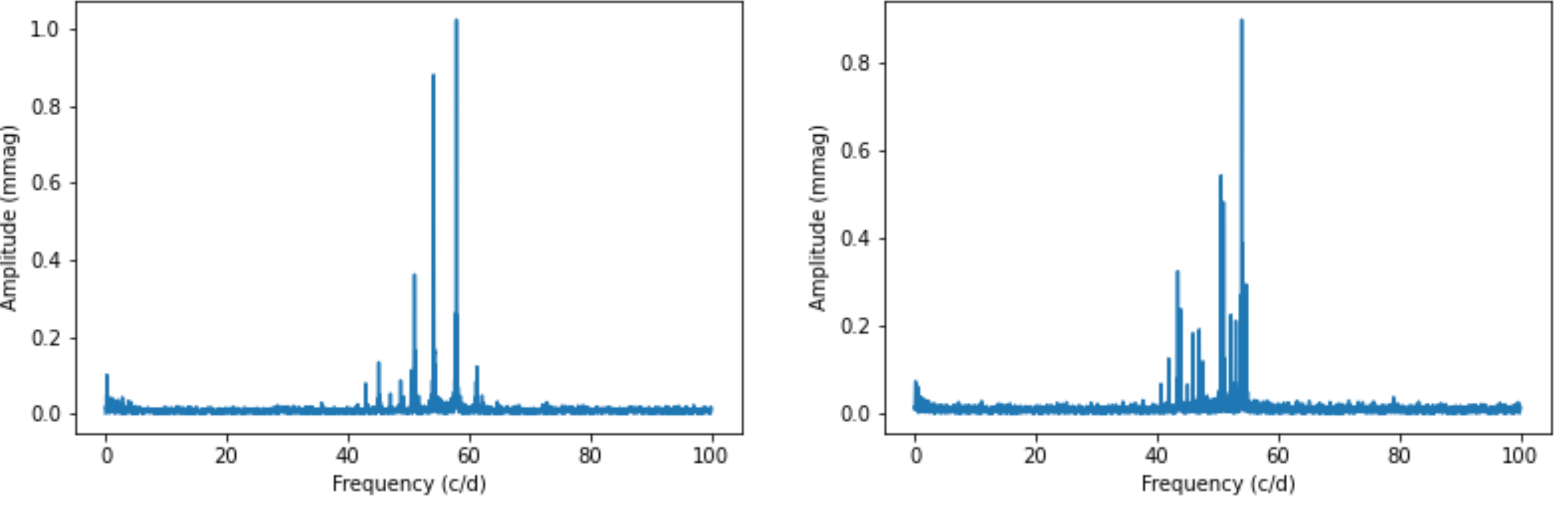}
    \caption{Periodograms for TIC 354792288 and TIC 285935852, showing very grouped frequencies between 40 and 60 $\mu Hz$}
    \label{fig:TIC88andTIC52periods}
\end{figure*}

TIC 354792288 and TIC 285935852 show a set of very grouped frequencies between 40 and 60 \muhz\ (Fig.~\ref{fig:TIC88andTIC52periods}), and don't show significant modes below $5 d^{-1}$, which indicates that they can probably be pure \dss, as it is defined in \citet{Grigahcene2010,Uytterhoeven2011}. With both stars, we have used $\nu$\textsubscript{max,0} for BF2018's and BK2018's relations, $\nu$\textsubscript{max,w} for BF2020's, and $\nu$\textsubscript{max,2D} for H2021's, in order to determine a mean effective temperature, $\widetilde{T}$\textsubscript{eff}, (see Tab.~\ref{tb:numax}), considering that the gravity darkening effect due to rotation makes the star hotter on the poles and cooler on the equator.  Regarding the uncertainties of the effective temperature, we have calculated them using the error propagation method, from the corresponding errors of $\nu$\textsubscript{max,0}, $\nu$\textsubscript{max,w} and $\nu$\textsubscript{max,2D} and the coefficients of their respective relations.\par

\footnotetext[3]{\href{https://pypi.org/project/echelle/}{Echelle 1.5.1}}

\begin{table}
	\centering
	\caption{Seismic indices of the selected targets from \ap\ to constrain the models, in order to determine its ages}
    \renewcommand{\arraystretch}{2.6}
	\addtolength{\tabcolsep}{2 pt}
	\resizebox{8.5cm}{!}{
    \begin{tabular}{cccccc}
			\hline
			TIC & $\fDnulow (\mu Hz)$ & $\omega_{r} (\mu Hz)$ & $\nu_{\text{max,0}} (\mu Hz)$ & $\nu_{\text{max,w}} (\mu Hz)$ & $\nu_{\text{max,2D}} (\mu Hz)$\\
			\hline
			410732825 & $63\pm1$ & $10\pm1$ & - & - & - \\
			354792288 & $83\pm1$ & - & $670.587\pm0.001$ & $624\pm15$ & $620\pm70$ \\
			285935852 & $82\pm1$ & $10\pm1$ & $625.026\pm0.001$ & $602\pm40$ & $605\pm70$  \\
			252829836 & $71\pm1$ & - & - & - & -  
    \end{tabular}
}
	\label{tb:indices}
\end{table}

\begin{table}
	\centering
	\caption{The different relations between the frequency at maximum power and the mean effective temperature for TIC 354792288 and TIC 285935852. References: $^{1}$\citeauthor{Forteza2018} \citeyear{Forteza2018},
	$^{2}$\citeauthor{Bowman2018} \citeyear{Bowman2018},
	$^{3}$\citeauthor{Forteza2020} \citeyear{Forteza2020},
	$^{4}$\citeauthor{Hasanzadeh2021} \citeyear{Hasanzadeh2021}}
    \renewcommand{\arraystretch}{2.6}
	\addtolength{\tabcolsep}{2.3 pt}
	\resizebox{8.5 cm}{!}{
    \begin{tabular}{cccc}
			\hline
			Relation $\nu_\mathrm{max} - \widetilde{T}_\mathrm{eff}$ &
			$\sigma (\%)$ & 
			TIC 354792288 $\widetilde{T}_\mathrm{eff}$ (K) & TIC 285935852 $\widetilde{T}_\mathrm{eff}$ (K)\\
			\hline
			$^{1}(2.39 \pm 0.20)\nu_\mathrm{max,0}$ (\muhz) + $(7110 \pm 50)$ & 5.87 & [8530:8898] & [8429:8779] \\
			$^{2}(22.7 \pm 4.0)\nu_\mathrm{max,0}(d^{-1})$ + $(6819 \pm 21)$ & - & [7881:8387] & [7808:8282] \\
			$^{3}(3.5 \pm 0.1)\nu_\mathrm{max,w}$ (\muhz) + $(6460 \pm 40)$ & 3.36 & [8477:8773] & [8327:8807] \\
			$^{4}(1.14 \pm 0.07)\nu_\mathrm{max,2D} + (1.22 \pm 0.01)$ (solar units) & - & [8107:8697] & [8076:8662] \\
    \end{tabular}
    }
	\label{tb:numax}
\end{table}

\section{The grid of models}\label{sec:Grid}

We have built a grid of 1D stellar models using the code \mesa\ for the evolution, and \filou\ for calculating the oscillation modes, because it takes into account the stellar distortion due to the centrifugal force in the oscillation frequency computation, and rotation up to second order in the perturbation approximation, including near-degeneracy effects. We have taken orders $2 \leq n \leq 8$ and degrees $0 \leq l \leq 2$, for calculating the low-order large separation of the models \citep{Suarez2014,RM2020}.\par
Regarding the evolutionary models, we built a grid using the parameters of Tab.~\ref{tb:grid}, delimiting ages between 20 and 200 Myrs. According to Z\textsubscript{base} values, we selected the Type2 tables for opacities. For the mixing length parameter we considered $\alpha = 2.0$, and a diffusion coefficient for mixing of elements of $D\_mix = 1/30$ \citep{Heger2000}. Overshooting was not considered. In total, 24965 models were calculated. 

\begin{table}
	\centering
	\caption{Selected values for the parameters of the grid of models built with MESA}
    \renewcommand{\arraystretch}{2}
	\addtolength{\tabcolsep}{8 pt}
	\resizebox{6 cm}{!}{
    \begin{tabular}{cccc}
			\hline
			Parameter & Range & Step\\
			\hline
			Age & [20, 200] Myr & 1 Myr\\
			M $(M_{\odot})$ & [1.6, 2.7] & 0.1 $M_{\odot}$\\
			$Z_{0}$ & [0.014, 0.020] & 0.002\\
			$\Omega / \Omega_{c}$ & [0.15, 0.25] & 0.05\\
			$\alpha$ & 2.0 & Fixed\\
    \end{tabular}
}
	\label{tb:grid}
\end{table}

\section{\texorpdfstring{Seismic determination of the age of $\boldsymbol{\alpha\ P\MakeLowercase{er}}$}{Seismic determination of the age of alpha Per}}\label{sec:Age}

\begin{figure*}   
    \centering
    \includegraphics[width=0.96\textwidth]{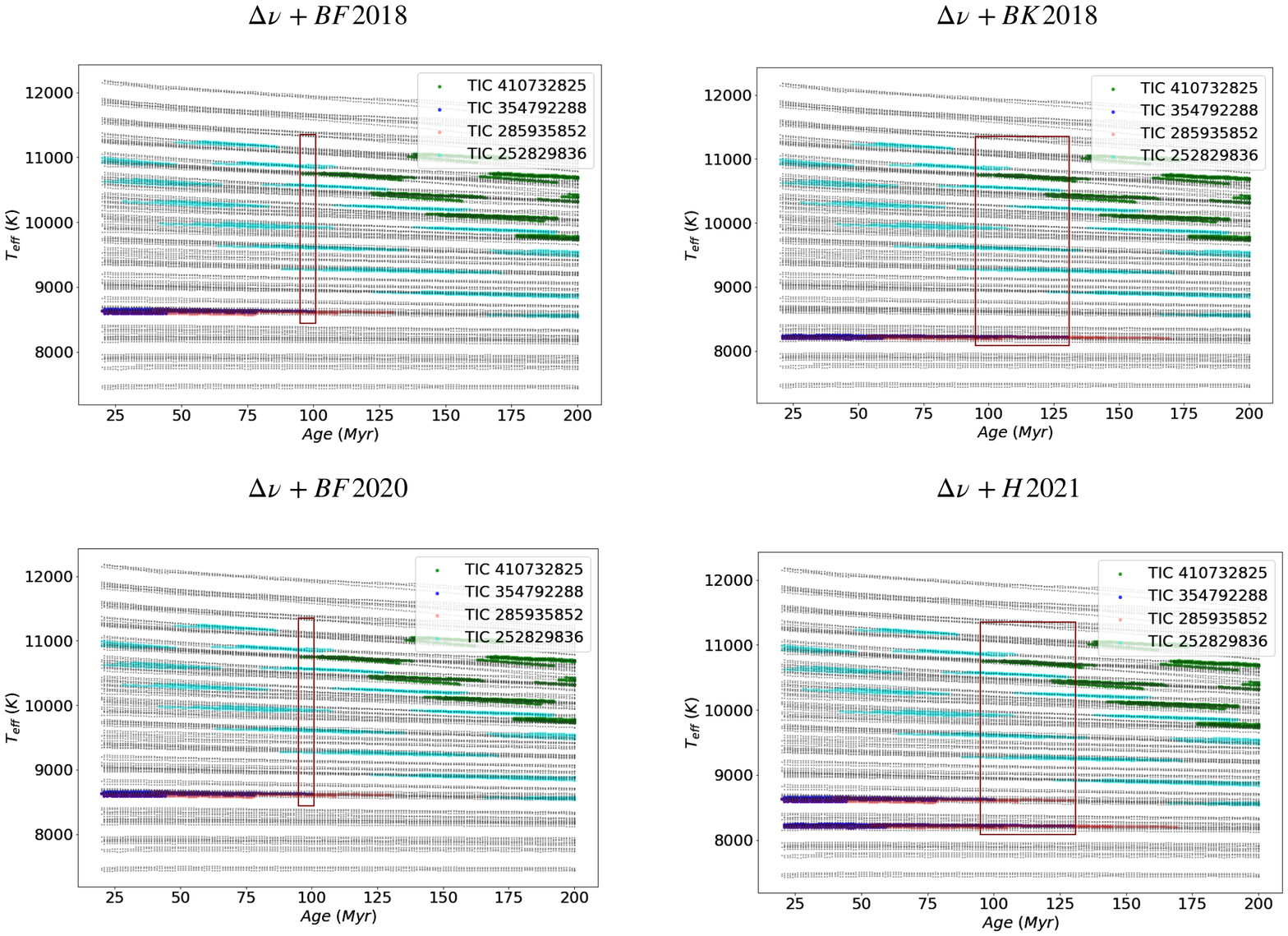}
    \caption{Constrained models for dating ages of the four stars, just using the low regime large separation ($\Delta \nu$), and also adding the different relationships between the frequency at maximum power and the effective temperature: $\Delta \nu +  BF2018$, $\Delta \nu + BK2018$, $\Delta \nu + BF2020$ and $\Delta \nu + H2021$}
    \label{fig:Ages_no_rot}
\end{figure*}

Fig.~\ref{fig:Ages_no_rot} shows the effective temperature vs age plots for all the calculated models, using the different relations between the frequency at maximum power and the effective temperature presented in Sec.~\ref{sec:Regularities}. The constrained models for TIC 410732825 are represented in green, for TIC 354792288 in blue, for TIC 285935852 in red and for TIC 252829836 in cyan.\par
We have compared the observed $\fDnulow$ value with the one calculated directly from the models. If we use the $\nu$\textsubscript{max,0}-T\textsubscript{eff} relation from BK2018 (top left panel) or $\nu$\textsubscript{max,2D}-T\textsubscript{eff} from H2021 (bottom left panel), then the constrained models for our four stars show common ages between 96 and 130 Myr. If we use the $\nu$\textsubscript{max,0}-T\textsubscript{eff} relation from BF2018 (top left panel) or the  $\nu$\textsubscript{max,w}-T\textsubscript{eff} relation from BF2020 (bottom left panel), the result is that the age of the cluster would be between 96 and 100 Myr, not too far from the most common age in the literature, around 90 Myr. \par
\begin{figure*}   
    \centering
    \includegraphics[width=0.96\textwidth]{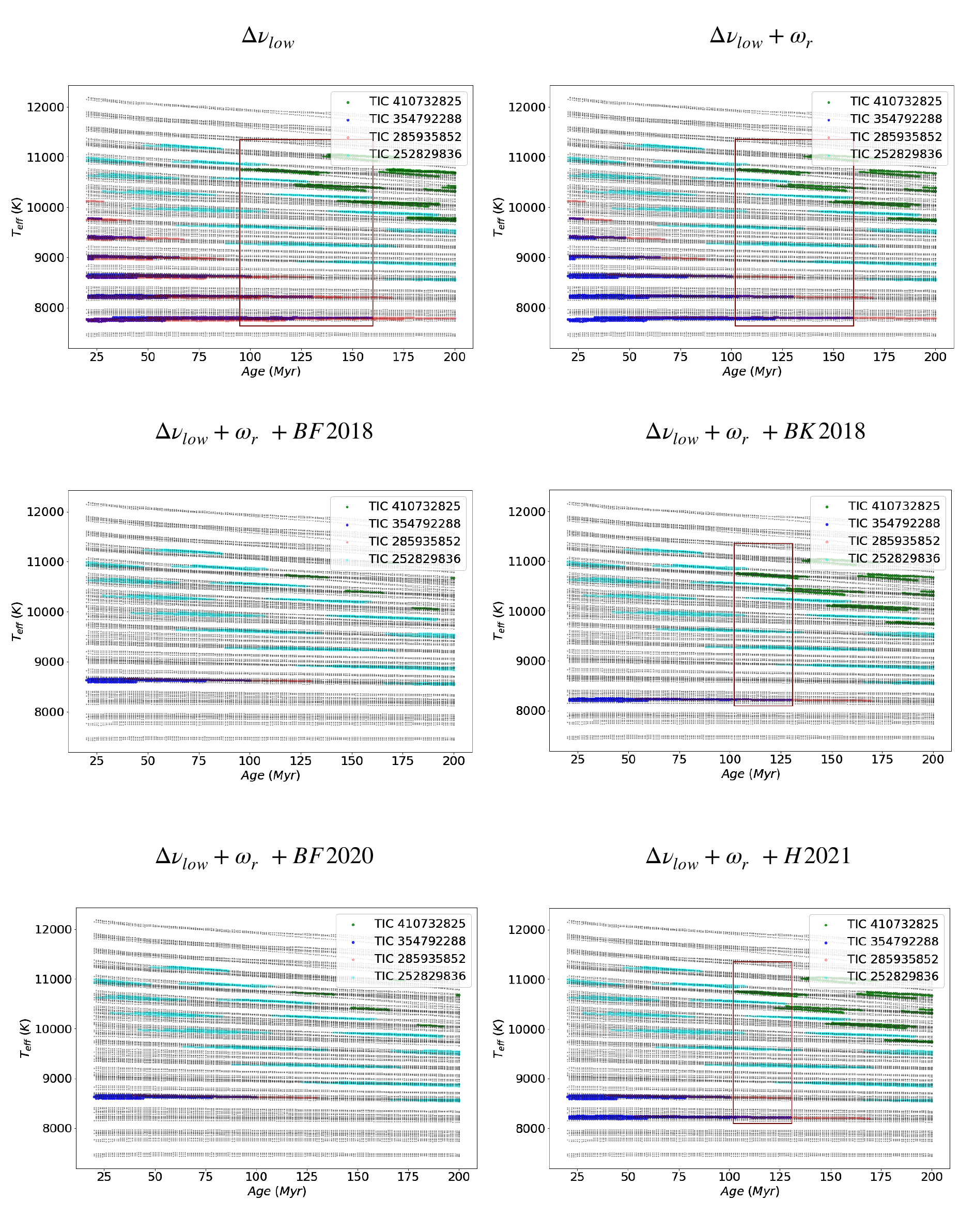}
    \caption{Constrained models for dating ages of the four stars, just using the low regime large separation ($\Delta \nu$), adding the angular rotation for TIC 410732825 and TIC 285935852 ($\Delta \nu + \omega_{r}$) and also adding the different relationships between the frequency at maximum power and the effective temperature: $\Delta \nu + \omega_{r} + BF2018$, $\Delta \nu + \omega_{r} + BK2018$, $\Delta \nu + \omega_{r} + BF2020$ and $\Delta \nu + \omega_{r} + H2021$}
    \label{fig:Ages_rot}
\end{figure*}

 
If we add as a constraint the rotation frequency obtained from the splittings in TIC 410732825 and TIC 285935852, then the scenario changes. In Fig.~\ref{fig:Ages_rot} we have plotted effective temperature vs age when using just the low-order large separation (top left panel), when  the rotation frequency of both stars are added to constrain the models (top right panel), and also when using the different relations between the frequency at maximum power and the effective temperature (middle and bottom panels). When we have used BK2018 (middle right panel) and H2021 (bottom right panel), two independent works, we obtained the same common ages, between 96 and 130 Myr. However, in the cases of BF2018 and BF2020, the result is that they don't show common ages.\par
Tab.~\ref{tb:parameters} shows the observed values of the main parameters of these four stars, and Tab.~\ref{tb:models} the values corresponding to the models when using as constraints the low-order large separation and the relation BF2020 (see Fig.~\ref{fig:Ages_no_rot}, bottom left panel). The agreement between the different relationships is greater when we do not constrain the models in rotation. Here we highlight the need to use tighter parameters in models with rotation, such as overshooting or the mixing length parameter, in order to improve the accuracy of the method. We have chosen BF2020 as reference because it was obtained with a quite large sample of pure \dss, as can be the case of the two stars with which we have used the relation. BK2018 and H2021 also used hybrid stars, which can impair the accuracy of the method. \par

Regarding TIC 410732825, the models tell us that it has a larger radius and a lower density than the ones calculated from the observed parameters. The density obtained from the large separation using the \citet{GH2017} relationship seems to corroborate this. The values of the projected rotation of \citet{Kounkel2019} and the seismic rotation are very similar, so the star seems to be equator on. This would explain why the possible values of the observed luminosity are below those of the models. If the observed parameters for TIC 252829836 were correct, then the star would have an age of hundreds of millions of years in order to be compatible with the measured large separation. This solution was therefore discarded, and points toward a wrong estimate of its physical parameters.\par

Our four stars, which seem to be members of the cluster according to \citet{Lodieu2019}, and do not seem to be binary systems, according to \citet{Kounkel2019}, point to an age for the cluster of at least 96 Myr. 

\begin{table*}
	\centering
	\caption{Parameters of the selected targets. References: $^{1}$\citeauthor{Stassun2019} \citeyear{Stassun2019},
	$^{2}$\citeauthor{Kounkel2019} \citeyear{Kounkel2019},
	$^{3}$\citeauthor{GH2017} \citeyear{GH2017}}
    \renewcommand{\arraystretch}{2.6}
	\addtolength{\tabcolsep}{2 pt}
	\resizebox{17cm}{!}{
    \begin{tabular}{ccccccccccc}
			\hline
			TIC & $M (M_{\odot})^{1}$ & $R (R_{\odot})^{1}$ & $\bar \rho (\bar \rho_{\odot})$ & $log g^{1}$ & T\textsubscript{eff}$(K)^{1}$ & $log(L/L_{\odot})^{1}$ & $vsin i(kms^{-1})^{2}$ & $\fDnulow (cd^{-1})$ & $\bar \rho_{\fDnulow}(\bar \rho_{\odot})^{3}$  \\
			\hline
			410732825 & [1.983:2.651] & [1.592:1.686] & [0.41:0.66] & [4.30:4.45] & [8851:9453] & [1.18:1.28] & [71:107] & [62:64] & [0.31:0.33] \\
			354792288 & [1.666:2.370] & [1.518:1.624] & [0.39:0.68] & [4.27:4.44] & [7849:8605] & [0.93:1.08] & [107:119] & [82:84] & [0.54:0.57]  \\
			285935852 & [1.465:2.063] & [1.545:1.645] & [0.33:0.56] & [4.20:4.36] & [7363:7815] & [0.84:0.92] & [68:73] & [81:83] & [0.53:0.56]  \\
			252829836 & [1.313:1.863] & [1.617:1.747] & [0.25:0.44] & [4.10:4.27] & [6994:7272] & [0.80:0.84] & [37:39] & [70:72] & [0.39:0.42] \\
    \end{tabular}
    }
	\label{tb:parameters}
\end{table*}

\begin{table*}
	\centering
	\caption{Constrained parameters of the models corresponding to our selected targets, using the relation BF2020}
    \renewcommand{\arraystretch}{2.6}
	\addtolength{\tabcolsep}{2 pt}
	\resizebox{16cm}{!}{
    \begin{tabular}{ccccccccccc}
			\hline
			TIC & $M (M_{\odot})$ & $R (R_{\odot})$ & $\bar \rho (\bar \rho_{\odot})$ & log g & T\textsubscript{eff}(K) & $log(L/L_{\odot})$ & $Z_{0}$ & $v(kms^{-1})$ & $\fDnulow (cd^{-1})$ & Age (Myr) \\
			\hline
			410732825 & [2.3:2.6] & [1.94:2.10] & [0.28:0.32] & [4.20:4.24] & [9731:11055] & [1.49:1.75] & [0.016:0.020] & [72:125] & [62:64] & [96:200]  \\
			354792288 & [1.75:1.85] & [1.51:1.54] & [0.49:0.52] & [4.32:4.34] & [8592:8666] & [1.05:1.06] & [0.016:0.020] & [71:116] & [82:84] & [20:100]  \\
			285935852 & [1.75:1.85] & [1.51:1.55] & [0.48:0.52] & [4.31:4.33] & [8579:8660] & [1.05:1.07] & [0.016:0.020] & [71:116] & [81:83] & [20:130]  \\
			252829836 & [1.9:2.6] & [1.68:1.94] & [0.36:0.40] & [4.24:4.30] & [8541:11248] & [1.13:1.71] & [0.016:0.020] & [69:126] & [70:72] & [20:200]  \\
    \end{tabular}
    }
	\label{tb:models}
\end{table*}

\section{Conclusions}\label{sec:Conclusions}
We have tested an asteroseismic method in order to determine the age of the young open cluster \ap. The technique is based on finding seismic indices, such as the low-order large separation, the frequency at maximum power and the rotation frequency in a sample of 11 \dss\ stars belonging to the field of the cluster. With them we have used the FT, the AC, the HFD and the ED. In four of them,  TIC 410732825, TIC 354792288, TIC 285935852 and TIC 252829836 we have measured the low-order regime large separation. With TIC 410732825 and TIC 285935852 we have found evidences of the angular rotation frequency. With the necessary caution, considering that they are near the ZAMS, we have used the relations between the frequency at maximum power and the effective temperature to constrain the age of \ap\ between 96 and 100 Myr.\par
The grid we have used can be extended to take into account the variability of parameters such as overshooting, the mixing length parameter or the diffusion of angular momentum.\par
The estimated age for \ap\ must be considered with caution until new studies with more \dss\ stars and/or a more detailed modeling allow us to confirm it. In any case, the method based on seismic indices looks like a promising tool for dating young stellar clusters. 
For the frequency analysis, we have developed a new code, \multim, (\textcolor{blue}{\href{https://github.com/davidpamos/MultiModes}{MultiModes}}), that we present in this paper. The code shows a good reliability after testing it with synthetic light curves, and after being compared with one of the most reliable codes in the field, \sigspec. Moreover, \multim\ is more versatile in terms of frequency analysis control. It has allowed us to find nine previously unknown \dss\ from the original sample of 32 analysed stars.\par 

\section*{Acknowledgements}

This publication is part of the project "Contribution of the UGR to the PLATO2.0 space mission. Phases C / D-1", funded by MCNI/AEI/PID2019-107061GB-C64.\par
DPO acknowledges MNRAS for the opportunity to publish his first scientific article, in particular to the assistant editor, Bella Lock, for her comments. He also appreciates the comments of the anonymous reviewer of this article. They have undoubtedly contributed to improving it. He also acknowledges all the co-authors of this work, especially to JCS and AGH, his thesis project directors, because without them, he would not have been able to publish this work. And he also appreciates the patience of his wife, Pilar, and his son, Fernando, for all the time they have allowed him to get to this point. \par
AGH acknowledges funding support from Spanish public funds for research under project PID2019-107061GB-C64 by the Spanish Ministry of Science and Education, and from ‘European Regional Development Fund/Junta de Andaluc\'{\i}a-Consejer\'{\i}a de Econom\'{\i}a y Conocimiento’ under project E-FQM-041-UGR18 by Universidad de Granada.\par
JPG acknowledges funding support from Spanish public funds for research from project PID2019-107061GB-C63 from the "Programas Estatales de Generaci\'on de Conocimiento y Fortalecimiento Cient\'ifico y Tecnol\'ogico del Sistema de I+D+i y de I+D+i Orientada a los Retos de la Sociedad", as well as from the State Agency for Research of the Spanish MCIU through the “Center of Excellence Severo Ochoa” award to the Instituto de Astrofísica de Andalucía (SEV-2017-0709).\par
SBF received financial support from the Spanish State Research Agency (AEI) Projects No. PID2019-107061GB-C64. He also thanks the resources received from the PLATO project collaboration with Centro de Astrobiología (PID2019-107061GB-C61).\par
This paper includes data collected with the TESS mission, obtained from the MAST data archive at the Space Telescope Science Institute (STScI). Funding for the TESS mission is provided by the NASA Explorer Program. STScI is operated by the Association of Universities for Research in Astronomy, Inc., under NASA contract NAS 5–26555.

\section*{Data Availability}

Tables of the most significant peaks, corresponding to each of the 11 \dss\ of our sample, extracted with MM, are available in \textcolor{blue}{\href{https://vizier.u-strasbg.fr/viz-bin/VizieR}{VizieR DataBase}}. In them we can find the values of the frequencies, amplitudes, phases, corresponding errors and S/N of the extracted peaks.

\section*{List of acronyms}
\begin{itemize}
    \item AC: Autocorrelation Function
    \item BF2018: \citet{Forteza2018}
    \item BF2020: \citet{Forteza2020}
    \item BK2018: \citet{Bowman2018}
    \item DFT: Discrete Fourier Transform
    \item ED: Èchelle Diagram
    \item HFD: Histogram of frequency differences
    \item FT: Fourier Transform
    \item H2021: \citet{Hasanzadeh2021}
    \item MM: MultiModes code
    \item SS: SigSpec code
\end{itemize}



\bibliographystyle{mnras}
\bibliography{bibl} 




\begin{appendices}
\section{Comparative analysis of efficiency between MM and SS}\label{appendixA}
\setcounter{figure}{0} 
\renewcommand{\thefigure}{A.\arabic{figure}}
Fig.~\ref{figa1} represents the executing time of both codes using the light curves of  11 \dss\ stars from our sample. MM is a factor two faster than SS in the cases where the number of extracted peaks is above 200 approximately. 
\begin{figure*}   
    \centering
    \includegraphics[width=0.7 \textwidth]{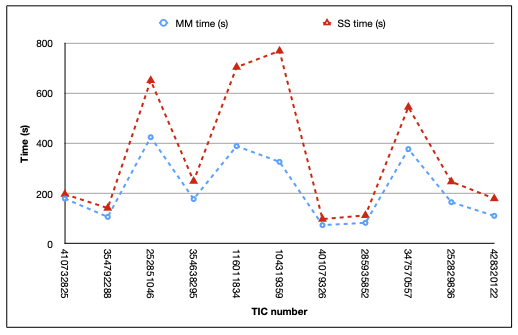}
    \caption{Comparative analysis of efficiency between MM and SS, in terms of computing time, done with the sample of 11 \dss\ stars}
    \label{figa1}
\end{figure*}

\section{Diagrams of regularities of TIC 285935852 and TIC 252829836}
\label{appendixB}

\begin{figure*}   
    \centering
    \includegraphics[width=0.9 \textwidth]{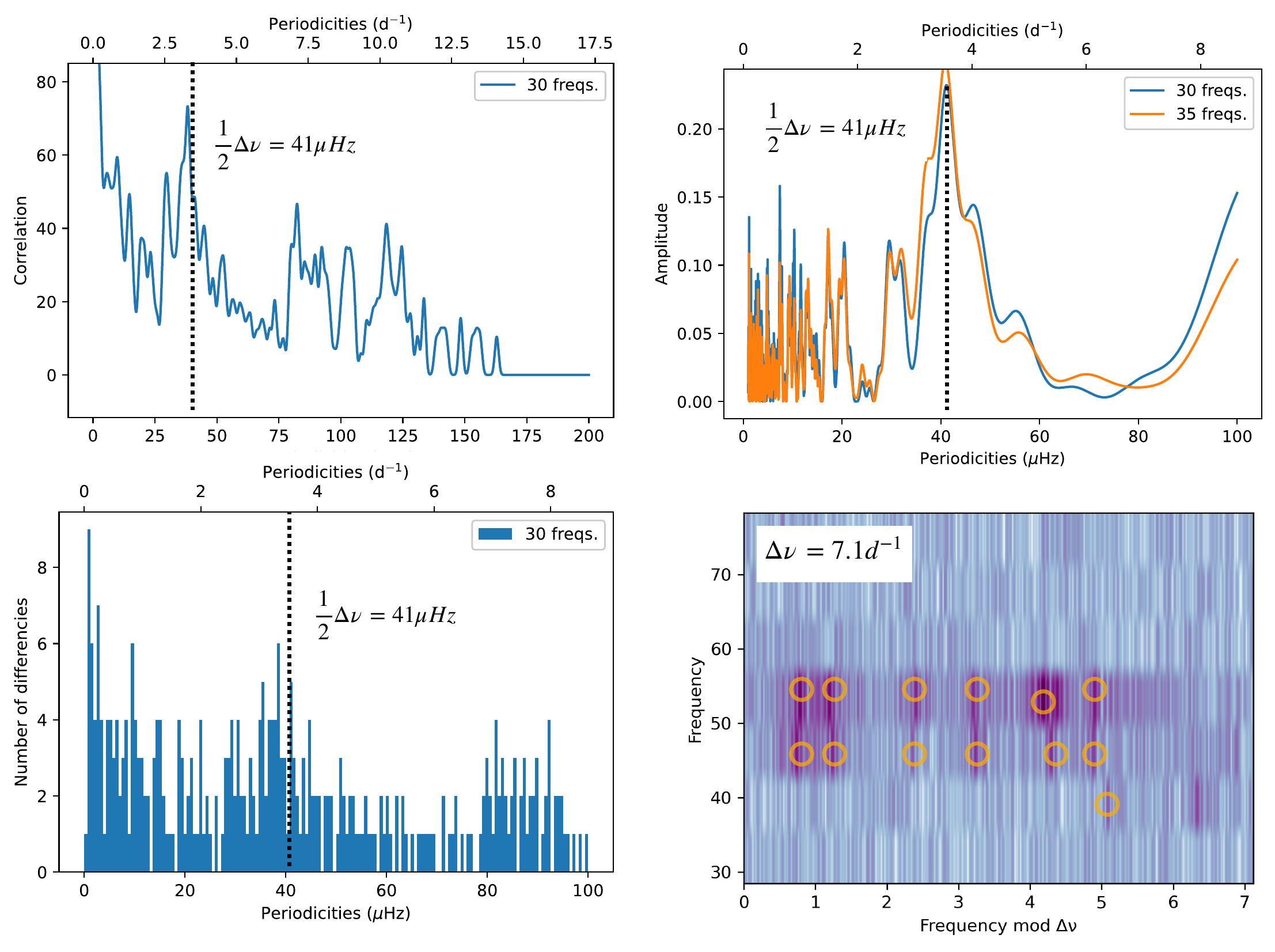}
    \caption{Measured regularities for TIC 285935852}
    \label{figa2}
\end{figure*}

\begin{figure*}   
    \centering
    \includegraphics[width=0.9 \textwidth]{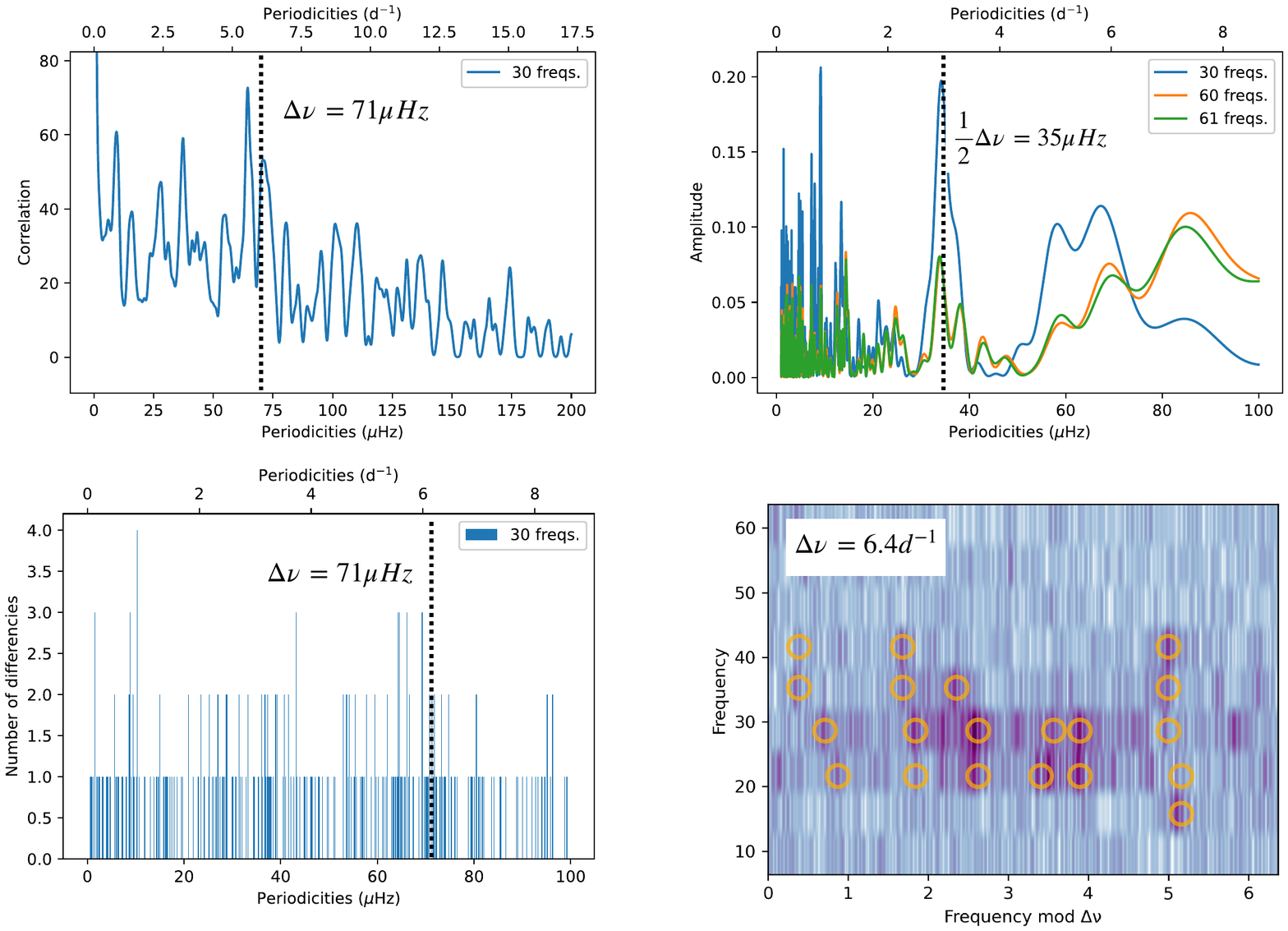}
    \caption{Measured regularities for TIC 252829836}
    \label{figa3}
\end{figure*}

We include here the AC, FT, HFD and ED of TIC 285935852 (Fig.~\ref{figa2}) and TIC 252829836 (Fig.~\ref{figa3}), which together with TIC 410732825  (Fig.~\ref{fig:TIC25}) and TIC 354792288 (Fig.~\ref{fig:TIC88}), complete the four stars analyzed for which regularities have been found in their corresponding frequency spectra.

\end{appendices}


\bsp	
\label{lastpage}
\end{document}

%% file: personal_set.tex
 \usepackage{multirow}

\newcommand{\dss}{$\delta$~Sct stars}

\newcommand{\corot}{{CoRoT}}

\newcommand{\filou}{{\sc{filou}}}

\newcommand{\sigspec}{{\sc SigSpec}}

\newcommand{\kepler}{{\it{Kepler}}}

\newcommand{\kms}{$\mathrm{km}~\mathrm{s}^{-1}$}

\newcommand{\muhz}{$\mu\mbox{Hz}$}

\newcommand{\cd}{$\mbox{d}^{-1}$}

\newcommand{\Dnulow}{$\Delta\nu_\mathrm{low}$}
\newcommand{\fDnulow}{\Delta\nu_\mathrm{low}}